\documentclass[twocolumn,superscriptaddress,amsfonts,amsmath,amssymb,prl,nobibnotes,showpacs]{revtex4-1}
\usepackage[colorlinks=true]{hyperref}
\usepackage{times}
\usepackage{graphicx}
\usepackage{subfig}
\usepackage{xspace}
\usepackage{color}
\newcommand{\LFT}{(LaFeO$_3$)$_2$/LaTiO$_3$\xspace}

\begin{document}
\title{Engineering charge ordering into multiferroicity}
\author{Xu He}
\affiliation{Beijing National Laboratory for Condensed Matter Physics, Institute of Physics, Chinese Academy of Sciences, Beijing 100190, China}


\author{Kui-juan Jin}
\email[Correspondence and requests for materials should be addressed to Kui-juan Jin. Email: ]{kjjin@iphy.ac.cn}
\affiliation{Beijing National Laboratory for Condensed Matter Physics, Institute of Physics, Chinese Academy of Sciences, Beijing 100190, China}
\affiliation{Collaborative Innovation Center of Quantum Matter, Beijing 100190, China}
\begin{abstract}
Multiferroic materials have attracted great interests but are rare in nature. In many transitional metal oxides, charge ordering and magnetic ordering coexist, so that a method of engineering charge-ordered materials into ferroelectric materials would lead to a large class of multiferroic materials. We propose a strategy for designing new ferroelectric or even multiferroic materials by inserting a spacing layer into each two layers of charge-ordered materials and artificially making a superlattice. One example of the model demonstrated here is the perovskite \LFT (111) superlattice, in which the LaTiO$_3$ layer acts as the donor and the spacing layer, and the LaFeO$_3$ layer is half doped and performs charge ordering. The collaboration of the charge ordering and the spacing layer breaks the space inversion symmetry, resulting in a large ferroelectric polarization. As the charge ordering also leads to a ferrimagnetic structure, the \LFT is multiferroic. It is expected that this work can encourage the designing and experimentally implementation of a large class of multiferroic structures with novel properties.
\end{abstract}
\maketitle

Multiferroic materials have attracted great interests, but there is few magnetic ferroelectric~\cite{nicola2000why}. Recent technical advances in the atomic-scale synthesis of oxides make it possible to artificially design~\cite{Mannhart26032010,hwang2012emergent,chakhalian2014colloquium} composite structures, which paves the way for engineering materials to get novel properties. Thus, strategies of designing multiferroic structures can be developed. For instance, hybrid improper ferroelectrics~\cite{bousquet2008improper,PhysRevLett.106.107204,ADMA:ADMA201104674,ADFM:ADFM201300210} with ABO$_3$/A'BO$_3$ ordered superlattice structures were designed by engineering the perovskite ABO$_3$. In these materials, the antiferroelectric rotation mode of B-O octahedra and the A-site ordering are brought together to form cooperative polar distortion.

Though most of ferroelectrics are induced by geometric distortion, there is a group of materials in which ferroelectricity is induced by charge ordering, as observed in perovskite manganites (PrCa)MnO$_3$~\cite{efremov2004bond}, magnetite Fe$_3$O$_4$~\cite{MIYAMOTO199451}, quasi-one-dimensional organics~\cite{PhysRevLett.86.4080}, the frustrated charge-ordered LuFe$_2$O$_4$~\cite{doi:10.1143/JPSJ.69.1526}, and complex manganites RMn$_2$O$_5$~\cite{hur2004electric}. Meanwhile, many materials with charge ordering are non-ferroelectric. Here we propose a method to design ferroelectric structures (or multiferroic structures if ferromagnetism is already exhibited in them) by making use of the charge ordering properties in these materials. The designing rule is to insert a spacing layer into each two layers of these charge-ordered materials and to make the structure as superlattice. Due to the multi-valence nature of transitional metal elements, transitional metal oxides with charge ordering are abundant, and many of them are magnetic. Thus, by modifying them into ferroelectrics, a large class of multiferroics can be expected.

\begin{figure}[h]
  \centering
  \includegraphics[width=0.47\textwidth]{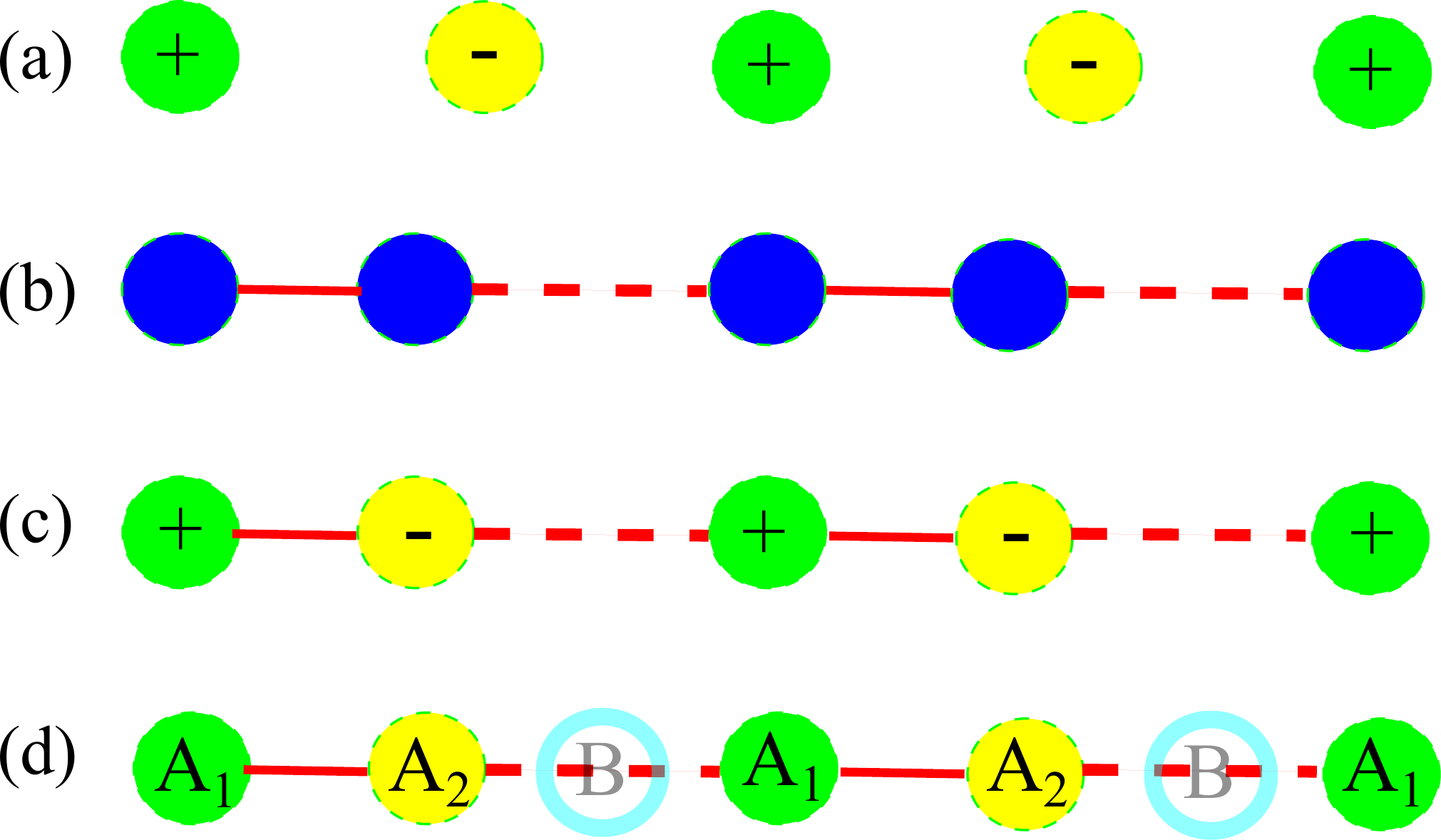}
  \caption{The scheme of charge-ordering-induced ferroelectricity in 1D chain. Green and yellow spheres denote two charge states. The solid and dashed lines denotes two kinds of bonds. (a) Site-centered charge ordering. (b) Bond-centered charge ordering. (c) Site- and bond-centered charge ordering. (d) The structure proposed in the present work, in which A$_1$ and A$_2$ are two layers of material A with different valence states, B is the spacing layer.}
  \label{fig:1dchain}
\end{figure}

To demonstrate the mechanism of ferroelectricity induced by charge ordering, we start from a simple one dimensional chain. There are two kinds of charge ordering, namely the site-centered charge ordering and the bond-centered charge ordering as shown in Figs.~\ref{fig:1dchain} (a) and (b), respectively. The former is that the electrons distribute alternately among sites so the charges on the neighboring sites are inequivalent; the latter is that the neighboring bonds become inequivalent. Neither of the two kinds alone leads to ferroelectricity as the space inversion symmetry is preserved, but the combination of both of them leads to a ferroelectric polarization, as shown in Fig.~\ref{fig:1dchain} (c). The theories and experimental results on charge-ordering-induced ferroelectricity are reviewed in Refs.~\cite{Jeroen2008multiferroicityCO,Kunihiko2014ferroCO}.

In the present work, the main point is to establish a method of designing superlattices by introducing the bond-centered charge ordering to the site-centered charge-ordered materials so that they become ferroelectric. The scheme of our design is shown in Fig.~\ref{fig:1dchain} (d). The charge ordering due to the multi-valence property of the transitional metal elements is mostly site-centered. Layers of charge-ordered material A with two valence states denoted as A$_1$ and A$_2$ are aligned alternately, forming an A$_1$A$_2$A$_1$A$_2$ pattern. Our designing principle is simply using the analogy to the 1D chain: the space inversion symmetry needs to be broken by two kinds of bonds between A$_1$ and A$_2$. By inserting a spacing layer of B into each two A layers, the pattern becomes A$_1$A$_2$BA$_1$A$_2$B as shown in Fig.~\ref{fig:1dchain} (d). If the A$_2$-B-A$_1$ is viewed as a kind of bond between A$_1$ and A$_2$, the two kinds of bonds of A$_1$-A$_2$ and A$_2$-B-A$_1$ are ordered. Thus, both site-centered charge ordering and bond-centered charge ordering exist in the (A)$_2$/B superlattice, leading to the ferroelectricity.
\begin{figure}[h]
  \centering
  \includegraphics[width=0.48\textwidth]{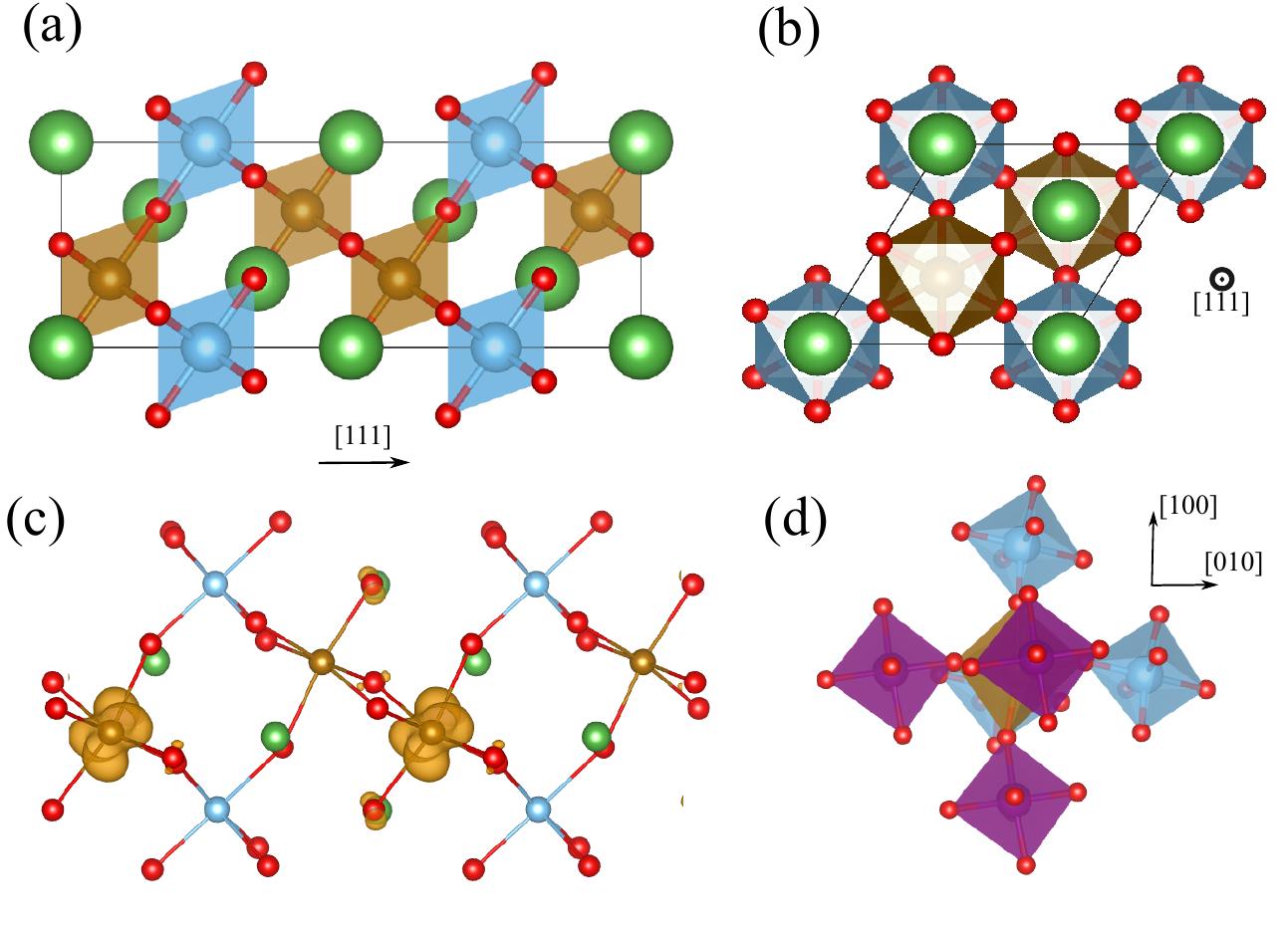}
  \caption{ (a) The structure of the \LFT superlattice. The green, brown, blue, and red spheres represent the La, Fe, Ti, and O atoms, respectively. (b) The structure viewed along the c axis ([111] direction). In (a) and (b) the lattices shown are the non-distorted simple cubic. (c) The isosurface of the density of the electrons doped into LaFeO$_3$. (d) The structure viewed from the [100] direction, where the Fe atoms in two neighboring (111) layers are painted in brown and purple to be discriminated.}
  \label{fig:structure}
\end{figure}

For a proof of the concept, we use LaFeO$_3$ and LaTiO$_3$ as A and B, respectively, and stack them along the [111] direction. The structure is shown in Figs.~\ref{fig:structure} (a) and (b). Based on density functional calculations, we show the multiferroicity induced by charge ordering in \LFT superlattice.Charge ordering is absent in LaFeO$_3$ by nature, therefore doping is needed. We used the LaTiO$_3$ layer not only as spacing layer to break the space inversion symmetry, but also as electron donor layer, which dopes one electron to each two LaFeO$_3$ units.Thus half Fe ions in LaFeO$_3$ becomes Fe$^{2+}$, while the other half ions remains Fe$^{3+}$. We found that charge ordering is formed in the LaFeO$_3$ layers along the [111] direction. So the (111) superlattice was designed so that the analogy to the 1D case is valid. With the charge ordering and the spacing layer, the structure is ferroelectric. The charge ordering in the present structure also introduces the ferrimagnetism. Thus, the \LFT is multiferroic.


%

The first principle calculations were performed with local spin density approximation (LSDA)~\cite{perdew1981self} and projector augmented wave (PAW)~\cite{kresse1999ultrasoft} method as implemented in Vienna \emph{ab\ initio} Simulation Package~\cite{kresse1996efficient} (VASP). We used a plane-wave basis set with the energy cutoff of 500 eV and a $5\times5\times5$\ $\Gamma$ centered k-points to integrate the Brillouin zone. The electron configurations of La $5s^25p^65d^16s^2$, Fe $2p^63d^64s^2$, Ti $2s^22p^63d^4$, and O $2s^22p^6$ were used. A Hubbard-like correction~\cite{dudarev1998electron} with $U$(Fe)=4.8 eV and $U$(Ti)=3.0 eV is used to better describe the on-site electron-electron interaction in the transitional metal elements~\cite{kleibeuker2014electronic}. The structures are fully relaxed until the residual forces are below $10^{-3}$ eV/\AA.
  The ferroelectric polarization were calculated using the maximally localized wannier function~\cite{mostofi2008wannier90} (MLWF) method, which is equivalent to the Berry phase approach~\cite{resta2007theory, king1993theory}. The MLWFs were constructed using the Wannier90 program and the VASP interface to it~\cite{franchini2012maximally}. We also checked the value using the Berry phase method. Since the ferroelectricity in the structure is supposed to result from the transferring of the electrons, the MLWF method were used so that we can track the Wannier centers and gain some intuition on how the ferroelectric polarization is formed.

\begin{figure}[b]
  \centering
  \includegraphics[width=0.47\textwidth]{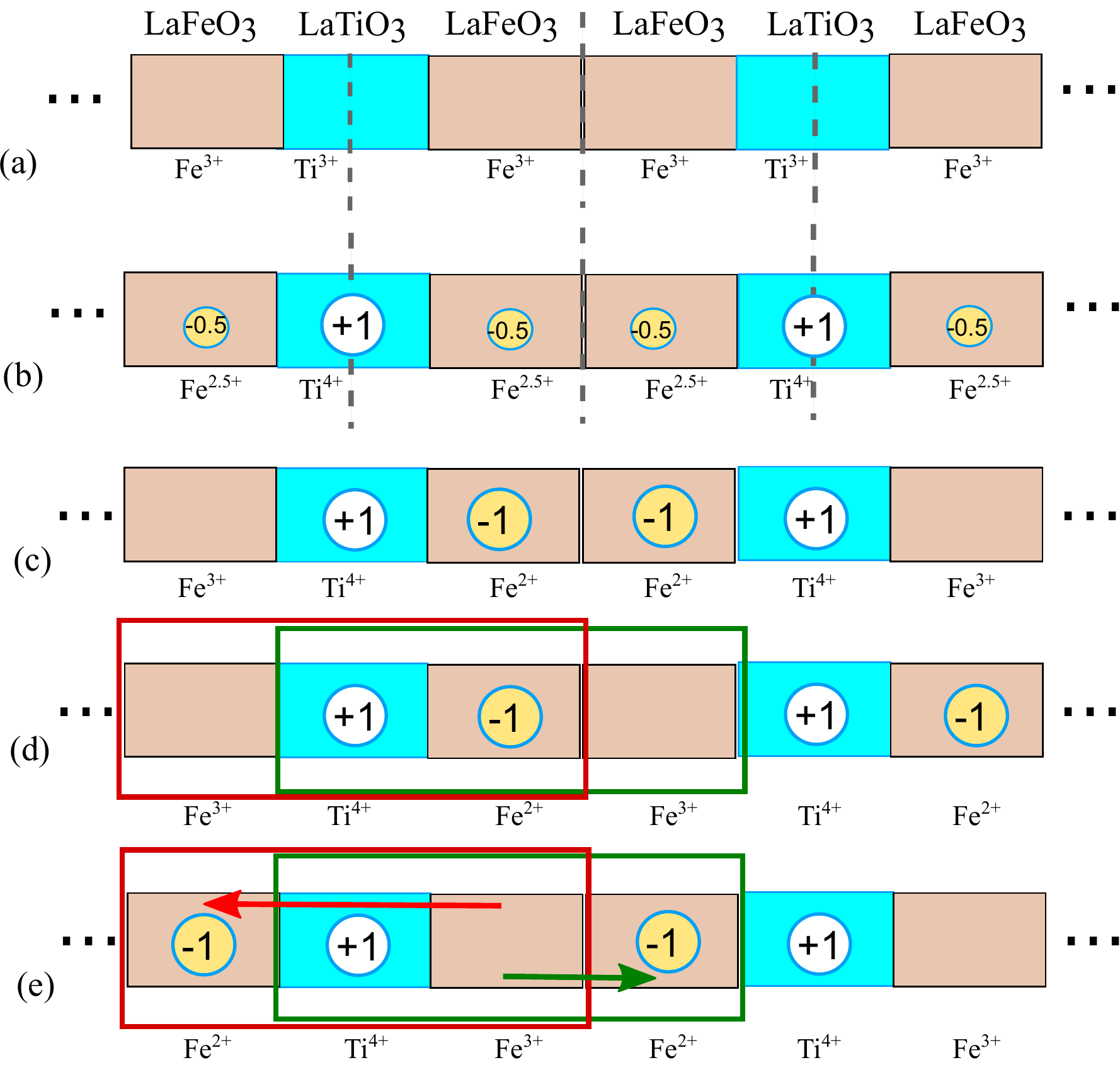}
  \caption{The scheme of the ferroelectricity in \LFT. (a) The structure without charge transfer between LaTiO$_3$ and LaFeO$_3$. (b) The structure with charge transfer but no charge disproportionation in LaFeO$_3$. (c) The structure with charge disproportionation and out-of-phase dipole alignment. (d) and (e) The structure with charge ordering and with opposite polarization direction. The dashed lines are the mirror planes. The green and red boxes denote two choices of unit cells for calculating the polarizations. The green and red arrows denote the corresponding paths of reversing the polarization.}
  \label{fig:fediag}
\end{figure}
The scheme of the ferroelectricity induced by the charge ordering in the \LFT (111) structure is shown in Fig.~\ref{fig:fediag}. The following conditions need to be satisfied to insure the structure being ferroelectric: (1) Electrons transfer from LaTiO$_3$ to LaFeO$_3$, otherwise all the Fe ions have the formal valence +3 (as shown in Fig.~\ref{fig:fediag} (a)). (2) The electron doped into LaFeO$_3$ only localizes in every other Fe ion, so the Fe ions have mixed valences of +2 and +3 (as shown in Fig.~\ref{fig:fediag} (d) or (e)), otherwise no charge ordering in LaFeO$_3$ can be developed (as in Fig.~\ref{fig:fediag} (b)). (3) The alignment of the Fe$^{2+}$-Fe$^{3+}$ should be in-phase like that shown in Fig.~\ref{fig:fediag} (d) or (e), otherwise no long-range ferroelectric domain can be formed (as in Fig.~\ref{fig:fediag} (c)). With all the above three conditions satisfied, the space inversion symmetry is broken and there is a macroscopic polarization in the structure as shown in Figs.~\ref{fig:fediag} (d) and (e). In the following text, we'll 
discuss the three conditions and the ferroelectric polarization in the (111) superlattice of \LFT in more detail.

\begin{figure}[htb]
  \centering
  \includegraphics[width=0.45\textwidth]{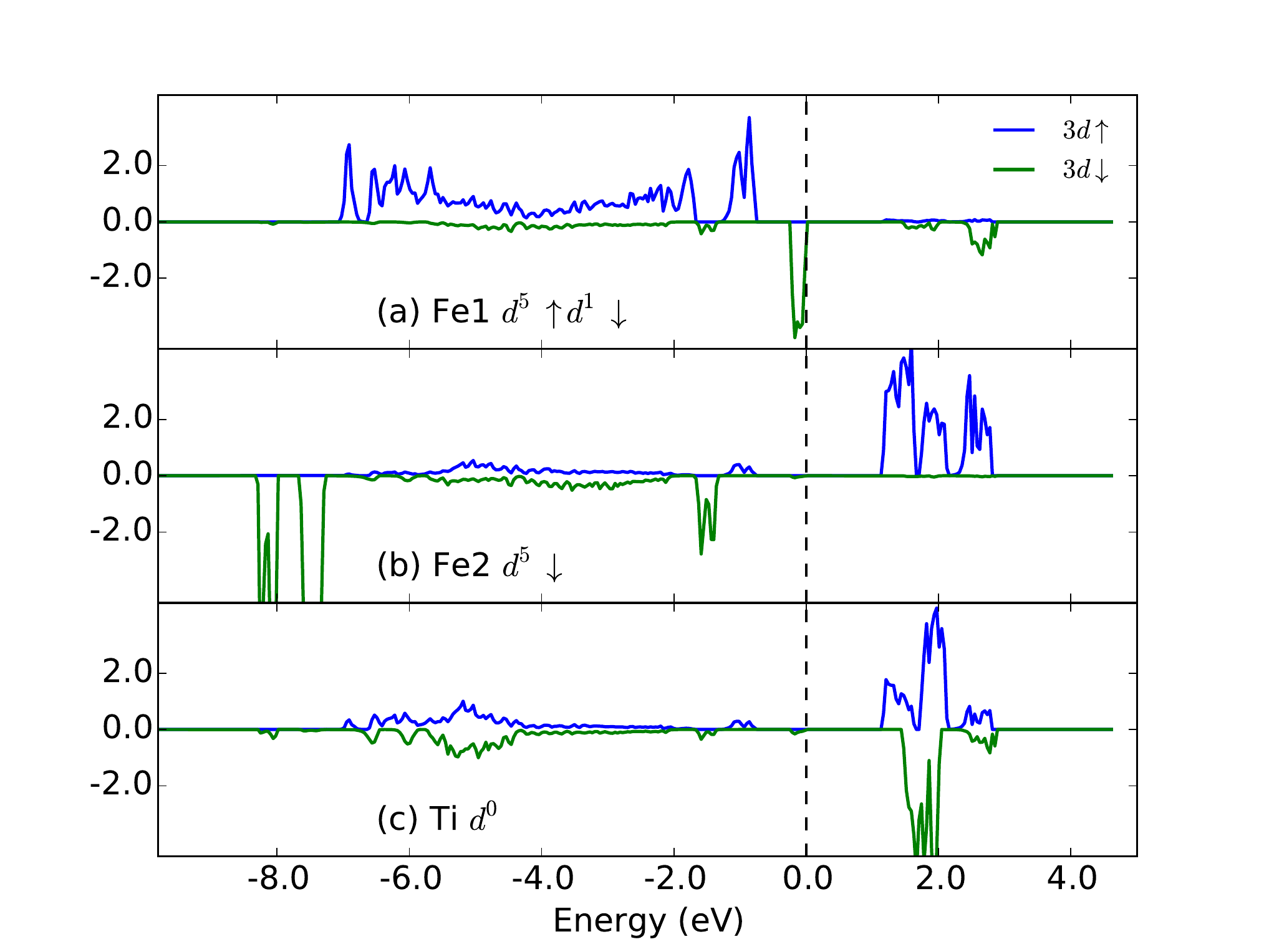}
  \caption{The density of the states projected on the (a) Fe$^{2+}$, (b) Fe$^{3+}$, and (c) Ti$^{4+}$ $3d$ orbitals. \label{fig:pdos} }
\end{figure}

Firstly, we discuss the electron transfer from the LaTiO$_3$ layers into LaFeO$_3$ layers. We used the LaTiO$_3$ as the electron donor because the $t_{2g}$ electrons of Ti$^{3+}$ tend to transfer to Fe$^{3+}$~\cite{kleibeuker2014electronic,PhysRevB.91.195145}. We found that each LaTiO$_3$ unit dopes about one electron into the LaFeO$_3$ in the \LFT (111) structure. The reconstruction of the Hubbard bands is the origin of the charge transfer, as Zhang \textit{et al.} suggested~\cite{PhysRevB.91.195145}. The bulk LaTiO$_3$ is a Mott-Hubbard insulator, in which each Ti ion has one electron on its $t_{2g}$ orbital. In bulk LaFeO$_3$, the Fe has a $d^5$ electronic configuration, and the conduction band minimum (CBM) is the unoccupied Fe $3d$ bands. The CBM of LaFeO$_3$ is higher than the occupied $t_{2g}$ band in LaTiO$_3$. However, if a Ti $t_{2g}$ electron transfer to the Fe site, the Fe 3$d$ bands would be reconstructed due to the on-site Coulomb interaction, so that the transferred electron to the Fe bands would have lower energy than the occupied Ti $t_{2g}$ bands. Therefore, the charge transfer is energetically favorable, resulting in the Ti $3d^0$ electronic configuration and half doping in Fe ions, as shown in the Fig.~\ref{fig:pdos}.

\begin{figure}[htb]
  \centering
  \includegraphics[width=0.45\textwidth]{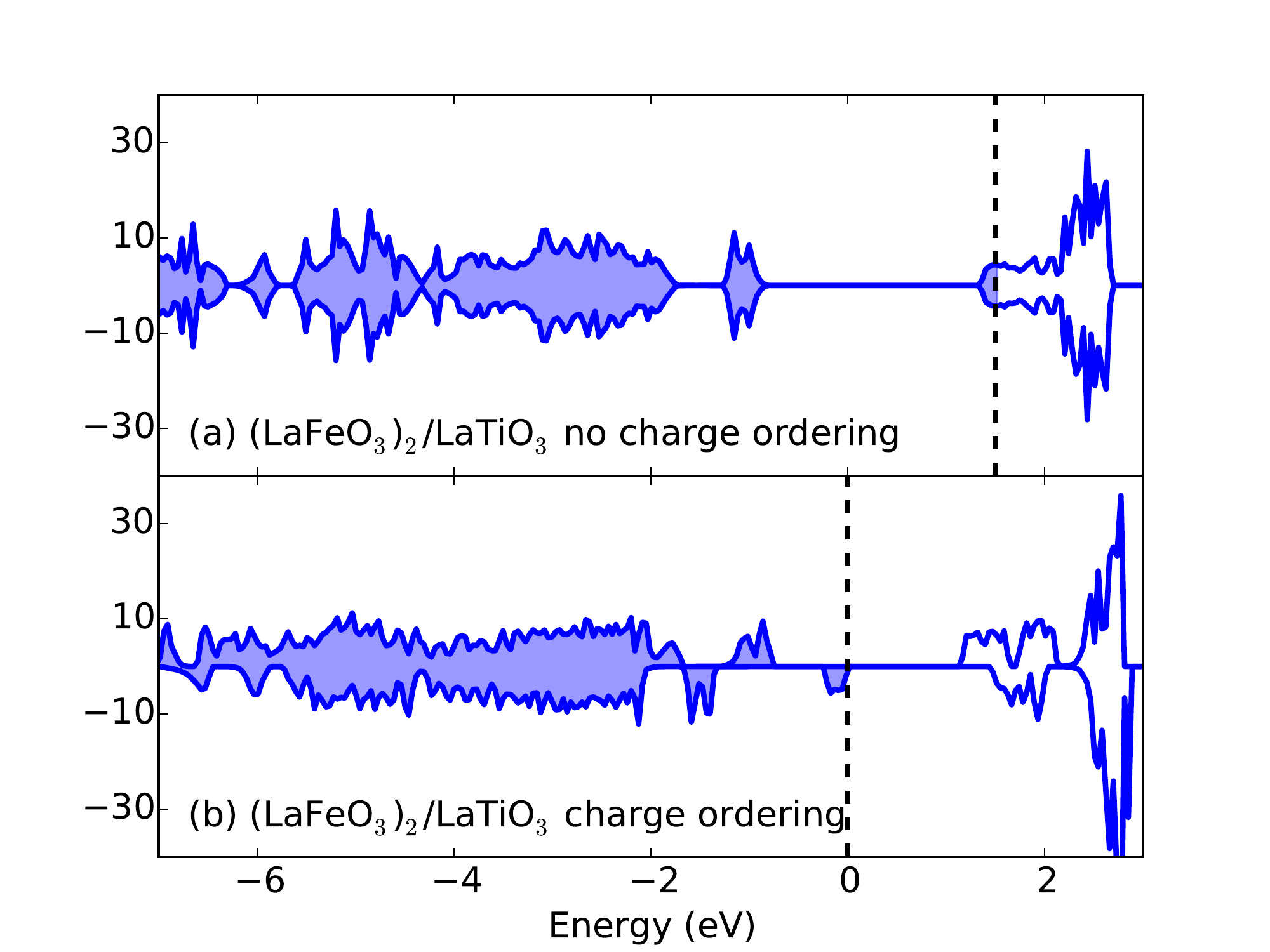}
  \caption{The total density of states for \LFT (a) without charge ordering and (b) with charge ordering. The positive and negative values are the spin up and spin down part, respectively.}
  \label{fig:tdos}
\end{figure}

 Secondly, we discuss the charge ordering in the nearest LaFeO$_3$ units. Each LaTiO$_3$ unit dopes about 1 electron into two LaFeO$_3$ units. The electron can either be shared by the two units equally (i.e. all Fe ions have the formal valence of +2.5) or be localized in one of them (i.e. Fe ions have mixed formal valences of +2 and +3). We compared the electronic structures without and with the charge ordering. The results of the density of states are shown in Fig.~\ref{fig:tdos}. Without mixed valence of the Fe ions developed in the LaFeO$_3$, the structure is metallic (Fig.~\ref{fig:tdos} (a)). Otherwise, the structure becomes insulating (Fig.~\ref{fig:tdos} (b)). We found that the structure with Fe site charge ordering is about 0.75 eV per formula unit (f.u.) lower in energy than that without charge ordering.

The localization of the doped electrons is because of the on-site Coulomb interaction of Fe electrons and the expansion of the Fe-O octahedra around the doped electrons staying on the Fe $t_{2g}$ orbitals. If the doped electron localizes on one Fe site, the on-site Coulomb interaction lowers the energy of the electron. (More details are in supplemental material Fig. S1.) Also, the expansion of the octahedron decreases the Coulomb energy of the doped electron as the distances of the electron on the Fe site to the negatively charged O anions are reduced.

The charge ordering modes in the LaFeO$_3$ layers sandwiched in LaTiO$_3$ layers were explored.( More detaisls are in supplemental material Fig. S2). The breathing mode charge ordering was found to be energetically favorable, in which the distribution of the doped electrons is shown in Fig.~\ref{fig:structure} (c). Thus the LaFeO$_3$ units in the same (111) plane are equivalent, and Fe$^{2+}$ and Fe$^{3+}$ align alternately along the [111] direction. All the neighboring Fe ions of the Fe$^{2+}$ are Fe$^{3+}$, and \textit{vice versa}, as shown in Fig.~\ref{fig:structure} (d). The breathing mode distortion of the Fe-O octahedra is the reason for the charge ordering in two LaFeO$_3$ layers connected to each other due to the octahedron-vertex sharing structure in perovskites. The volumes of the Fe$^{2+}$-O octahedron and Fe$^{3+}$-O octahedron are 11.2 \AA$^3$ and 10.0 \AA$^3$, respectively (details of the Fe-O bond lengths are in supplemental material Fig. S3).

Thirdly, we discuss the alignment of electric dipoles along the [111] direction. In each unit of (LaFeO$_3$)$_2$/LaTiO$_3$, there is an electric dipole of (LaTi$^{4+}$O$_3$)$^+$-(LaFe$^{2+}$O$_3$)$^-$. The dipoles in the chain along the superlattice can align in-phase or out-of-phase, which correspond to two kinds of Fe valance alignments, namely the Fe$^{3+}$-Ti$^{4+}$-Fe$^{2+}$-Fe$^{3+}$-Ti$^{4+}$-Fe$^{2+}$ alignment (Figs.~\ref{fig:fediag} (d)) and the Fe$^{3+}$-Ti$^{4+}$-Fe$^{2+}$-Fe$^{2+}$-Ti$^{4+}$-Fe$^{3+}$ alignment (Fig.~\ref{fig:fediag} (c)). The in-phase alignment leads to a ferroelectric domain; the out-of-phase alignment can be seen as the 180$^\circ$ wall between two polarization domains. We found that the structure cannot converge to the out-of-phase case, implying that it is unstable. One reason for the favoring of the in-phase structure is the dipole-dipole interaction. The energy is lowered if the dipoles are aligned in-phase. The other reason for the favoring of the in-phase alignment could be the elasticity of the LaTiO$_3$ layer. In the in-phase case, each Ti-O octahedron shares vertices with three Fe$^{2+}$-O octahedra and three Fe$^{3+}$-O octahedra. Whereas in the out-of-phase case, each of half the Ti-O octahedra shares vertices with six Fe$^{2+}$-O, and each of the other half shares vertices with six Fe$^{3+}$-O octahedra. Thus the Ti-O octahedra in the two cases have different sizes. The deviation of the sizes of the Ti-O octahedra from their optimal sizes costs elastic energy. If the energy cost for the in-phase case is smaller than that for the out-of-phase case, the structure with in-phase dipole alignment would be stabilized.

With the charge ordering along the [111] direction and the spacing layers, the space inversion symmetry in the structure of \LFT is broken, leading to a macroscopic polarization. Here we discuss the ferroelectric polarization in \LFT. The two possible polarization states as shown in Fig.~\ref{fig:fediag} (d) and (e) can be switched to each other with external field. In the modern theory of polarization, there is an ambiguity in the choice of unit cell when calculating the total polarization. Thus the allowed value of the polarization is not an unique one but a ‘lattice’ of values, which has the form of $\mathbf{P} = \mathbf{P_0} + n e \mathbf{R}/\Omega$, where $\mathbf{P_0}$ is a polarization value, $e$ is the unit electron charge, $\mathbf{R}$ is a lattice vector, $\Omega$ is the volume of the unit cell, and $n$ is an integer. The polarization quantum defined as $e\mathbf{R}/\Omega$ is 61.7 $\mu$C/cm$^2$ along the [111] direction in \LFT. The ferroelectric spontaneous polarization, which is the deviation of the polarization from that of the central symmetric structure, equals to half the difference between the polarization of the positively polarized state and that of the negatively polarized state ($\mathbf{P_s}= (\mathbf{P_+}-\mathbf{P_-})/2$). Thus the allowed values of the ferroelectric polarization are a ‘lattice’ of values with the interval of $e \mathbf{R}/2\Omega$. The calculated value of the polarization depends on the choice of the unit cell. In Figs.~\ref{fig:fediag} (d) and (e), the red and green boxes represent two choices, and correspondingly, the polarization states are reversed by moving an electron through the paths represented by the red and green arrows, respectively. Using the maximally localized wannier function methods, the calculated ferroelectric spontaneous polarizations are 21.4 $\mu$C/cm$^2$ and -9.5 $\mu$C/cm$^2$ (the minus sign means the opposite polarization direction) with the two choices, respectively. The difference of the two values is just  $e \mathbf{R}/2\Omega$ (30.9 $\mu$C/cm$^2$). To decide which path is taken is not a trivial task; Neaton \textit{et al.} proposed that multi paths can be taken, allowing for different ferroelectric polarization values~\cite{neaton2005first}. The allowed values of the ferroelectric polarization are (21.4 + 30.9 $n$) $\mu$C/cm$^2$, with $n$ being an integer.

The charge ordering also affects the magnetic structure in \LFT. We calculated the magnetic structure and found that the neighboring Fe-Fe magnetic interactions are antiferromagnetic, which is due to the superexchange of the $3d$ electrons of Fe ions~\cite{Fesuperexchange}. (More details are in supplemental material Fig. S4.) The antiferromagnetic interaction results in the alternating spin up and down alignment along the [111] direction and the parallel spin alignment in the (111) plane. Because of the charge ordering is also along the [111] direction, all Fe$^{2+}$ ions are in one (111) plane and are spin up, while all Fe$^{3+}$ ions are in the neighboring plane and are spin down, thus a ferrimagnetic structure is formed. The densities of states projected on 3$d$ orbitals of Fe$^{2+}$ and Fe$^{3+}$ as in Fig.~\ref{fig:pdos} show that the electronic configurations of the 3$d$ electrons of Fe$^{2+}$ and Fe$^{3+}$ ions are $d^5\uparrow d^1\downarrow$ and $d^5 \downarrow$, respectively. The Ti ions have $3d^0$ configurations and thus has 0 spin. The spin moments add up to 1 $\mu_B$/f.u., consistent with the caluculation result.

The synthesizing of the \LFT might be difficult. There have been only a few reports on the synthesizing of LaFe$_{1-x}$Ti$_{x}$O$_3$~\cite{Phokha2015118}. The good news is that the recent advances in the angstrom-scale layer-by-layer synthesis enables the fabricating of atomic-scale superlattices. The growth of both LaFeO$_3$ and LaTiO$_3$ monolayers has been reported~\cite{Ueda1064,disa2015orbital,cao2016engineered}. We also noted that (LaFeO$_3$)$_m$/(LaTiO$_3$)$_2$  heterostructures ($m$=2,4, 6, 8) has been synthesized recently, in which the Ti$^{4+}$ and Fe$^{2+}$ ions were found at the LaFeO$_3$/LaTiO$_3$ interface layers and Fe$^{3+}$ were found in the other layers~\cite{kleibeuker2014electronic}. We highly expect the synthesizing of the (111) superlattice \LFT.

  Many transitional metal oxide materials adopt charge ordering~\cite{Attfield2006861}, like vanadites~\cite{doi:10.1143/JPSJ.71.385}, manganites~\cite{PhysRev.100.564}, ferrites~\cite{PhysRevB.62.844}, cobalts~\cite{cobaltCO}, and nicklates~\cite{PhysRevLett.88.126402,PhysRevB.80.245105}, providing a lot of candidates for the charge-ordering layers. The magnetic interactions vary in these candidates, thus many kinds of magnetic properties with them are possible. In some of them, novel phenomenons related to the charge ordering were observed. For example, in colossal magnetic resistance material of manganites, the magnetic field can melt the charge ordering, which means that a magnetic control of the ferroelectric polarization could be realized if the material is engineered into a ferroelectric structure. There are abundant physics at the oxide interfaces~\cite{zubko2011interface}. Though not all of them are compatible with the charge-ordering-induced ferroelectricity, they  can bring novel phenomenons at the interfaces of the spacing layer and the charge-ordering layers. The charge ordering direction, depending on the distortion pattern of the lattice~\cite{PhysRevB.88.054101}, can also vary, so that various geometric structures can be designed. Thus, a large class of structures can be designed and some novel phenomenons are expected.

In the present work, we propose a method to engineer charge-ordered materials into ferroelectric materials or even multiferroic materials with magnetism being already presented. By inserting one spacing layer of B into each two layers of charge-ordered material A, (A)$_2$/B superlattice is formed, in which the space inversion symmetry is broken, leading to the ferroelectricity. We designed a \LFT (111) structure as a proof of the concept and investigate it by first principle calculations. Each LaTiO$_3$ unit dopes one electron into two LaFeO$_3$ units. The LaFeO$_3$ forms an alternating Fe$^{2+}$/Fe$^{3+}$ charge ordering along the [111] direction. With the LaTiO$_3$ layer inserted, the space inversion symmetry is broken and the structure becomes ferroelectric. The ferroelectric spontaneous polarization is about (21.4 + 30.9 $n$) $\mu$C/cm$^2$, where $n$ is an integer,  depending on the path of the polarization switching taken. The anti-parallel alignment of spins in Fe$^{2+}$ and Fe$^{3+}$ leads to a total net magnetic moment, so the structure is also ferrimagnetic. Further designing and experimental implementing of a large class of new multiferroic materials stimulated by this work are highly expected.
\begin{acknowledgments}
The work was supported by the National Basic Research Program of China (Nos. 2014CB921001 and 2012CB921403), the National Natural Science Foundation of China (No. 11134012), and the Strategic Priority Research Program (B) of the Chinese Academy of Sciences (No. XDB07030200).
\end{acknowledgments}
\bibliography{mybib.bib}

\pagebreak
\widetext
\begin{center}
\textbf{\large Supplemental Material for \\Engineering Charge Ordering into Multiferroicity }
\end{center}
\setcounter{equation}{0}
\setcounter{figure}{0}
\setcounter{table}{0}
\setcounter{page}{1}
\makeatletter
\renewcommand{\theequation}{S\arabic{equation}}
\renewcommand{\thefigure}{S\arabic{figure}}
\renewcommand{\bibnumfmt}[1]{[S#1]}
\renewcommand{\citenumfont}[1]{S#1}
\section{Dependence of Charge Ordering on $U$(F\lowercase{e})}
\begin{figure}[h]
  \centering
  \includegraphics[width=0.57\textwidth]{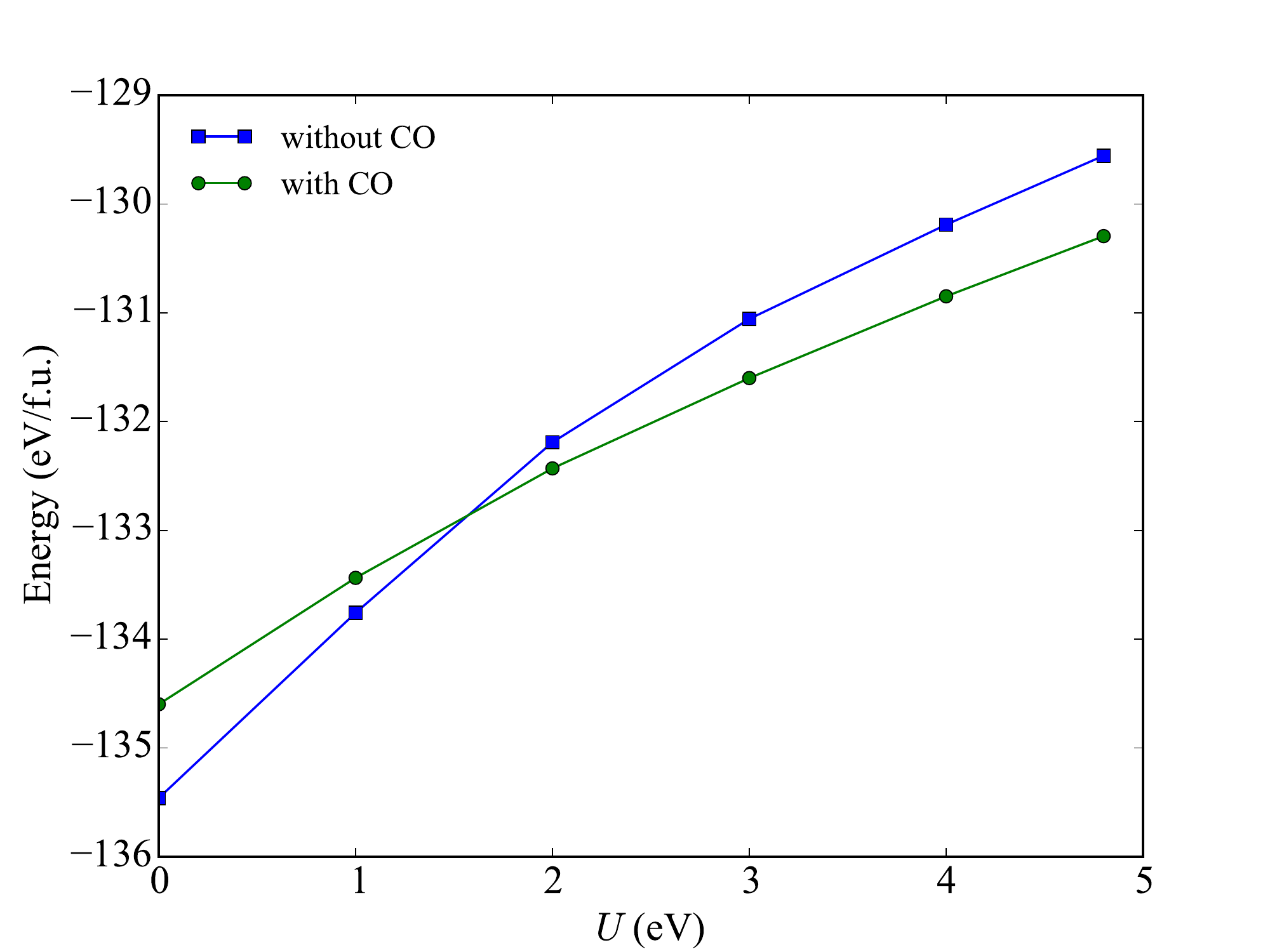}
  \caption{The energy dependence on $U$(Fe).}
  \label{fig:COnCO}
\end{figure}

The energies of the structure with and without charge ordering (CO) were calculated with various $U$(Fe). The results are shown in Fig. \ref{fig:COnCO}. The structure with Charge ordering is energetically favorable if $U$(Fe)$\geq$2 eV. The difference between the energy of the charge-ordered structure and that of the non-charge-ordered structure is larger with larger $U$(Fe), implying that the charge ordering is driven by the on-site electron-electron Coulomb interaction. The energy of the structure without charge ordering is calculated by keeping the space inversion symmetry while relaxing the other degrees of freedom. The structure with charge ordering is fully relaxed.
\newpage

\section{Charge Ordering Modes in L\lowercase{a}F\lowercase{e}O$_3$}
\begin{figure}[h]
  \centering
  \includegraphics[width=0.87\textwidth]{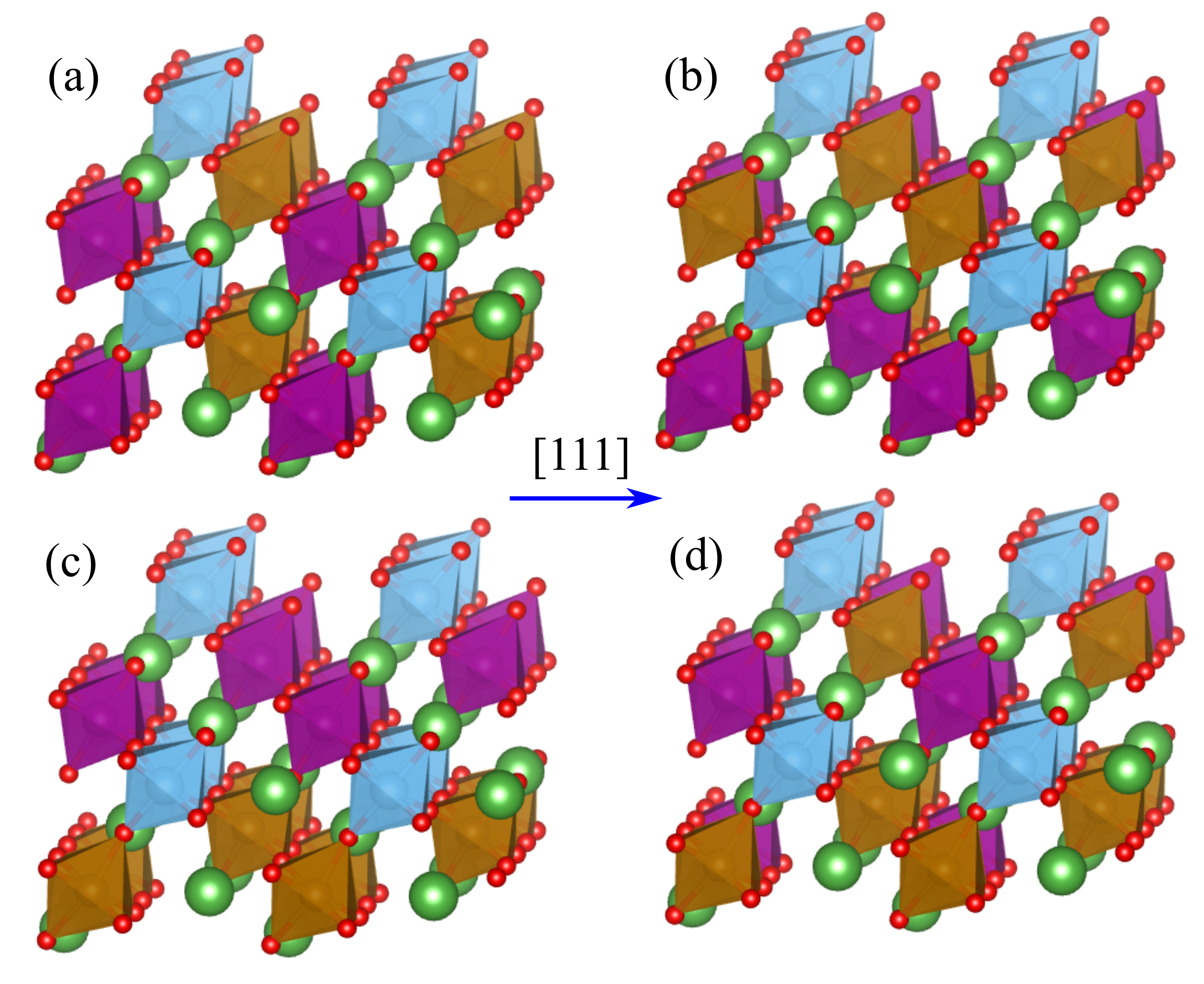}
  \caption{The scheme of the charge ordering modes.The green, blue, red spheres are the La, Ti, and O ions, respectively. The Fe$^{2+}$ and Fe$^{3+}$ ions are in brown and the purple respectively. The structure shown in (a) has the breathing mode charge ordering. }
  \label{fig:COmodes}
\end{figure}

 We compared the energies of four possible modes as shown in Fig. \ref{fig:COmodes}. The energies of the structures shown in Figs. \ref{fig:COmodes} (b), (c), and (d) are 45 meV/f.u., 83 meV/f.u., and 301 meV/f.u. larger than that of the structure in  Fig. \ref{fig:COmodes}(a), respectively.Therefore, the favored charge ordering mode is the breathing mode as shown in  Fig. \ref{fig:COmodes} (a). The structures were relaxed from the initial structures with Fe$^{2+}$-O bond lengths increased and Fe$^{3+}$-O bond lengths decreased.

\newpage
\section{The bond lengths}
\begin{figure}[h]
  \centering
  \includegraphics[width=0.67\textwidth]{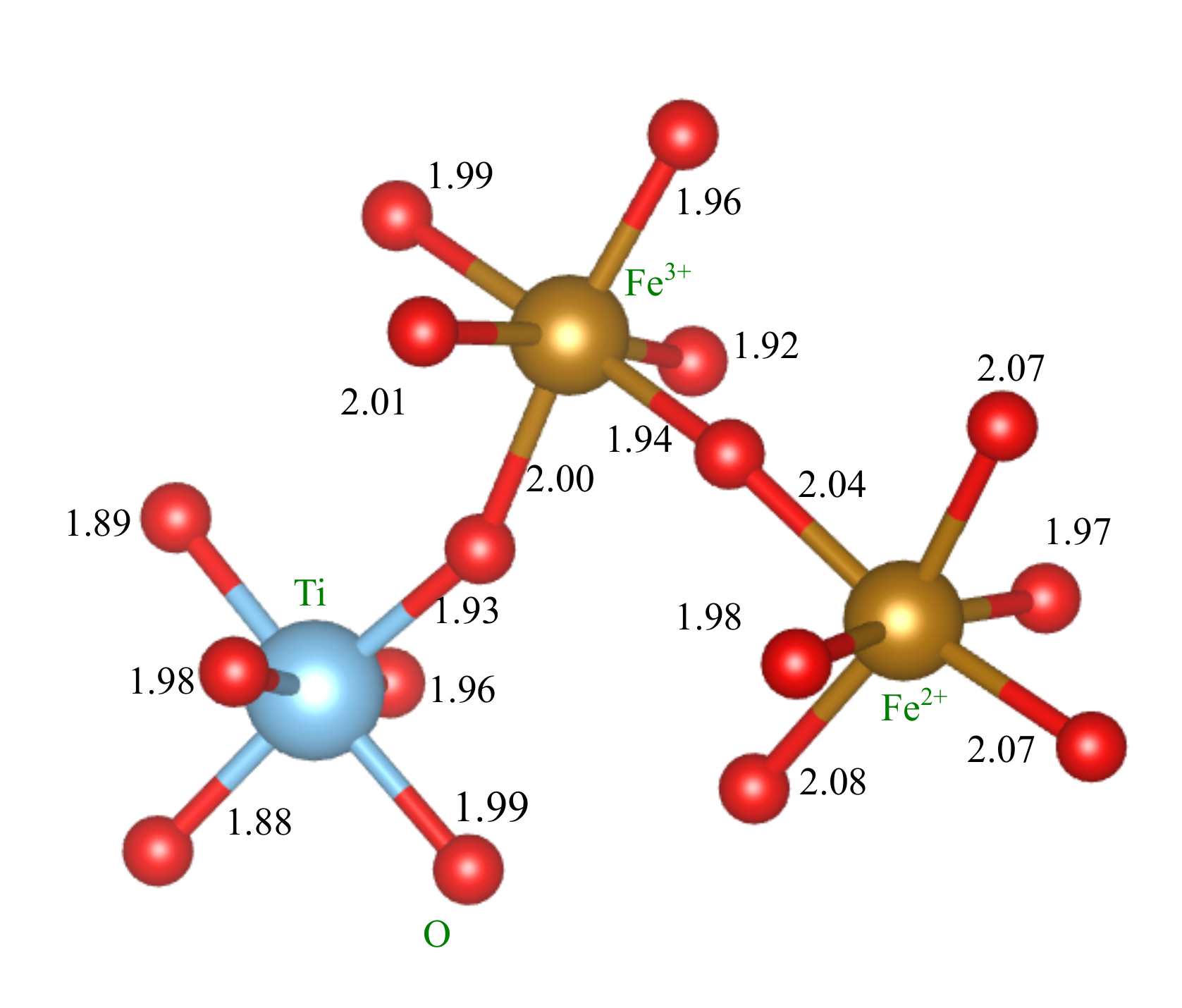}
  \caption{The bond lengths in the Fe-O and Ti-O octahedra. }
  \label{fig:bonds}
\end{figure}

The details of the Fe-O and Ti-O bonds in the breathing mode charge ordered structure are shown in Fig. \ref{fig:bonds}. The average Fe$^{2+}$ bond length is 2.03 \AA; the average Fe$^{3+}$ bond length is 1.97 \AA; The average Ti-O bond length is 1.94 \AA.
\newpage

\section{The magnetic structure}
\begin{figure}[h]
  \centering
  \includegraphics[width=0.57\textwidth]{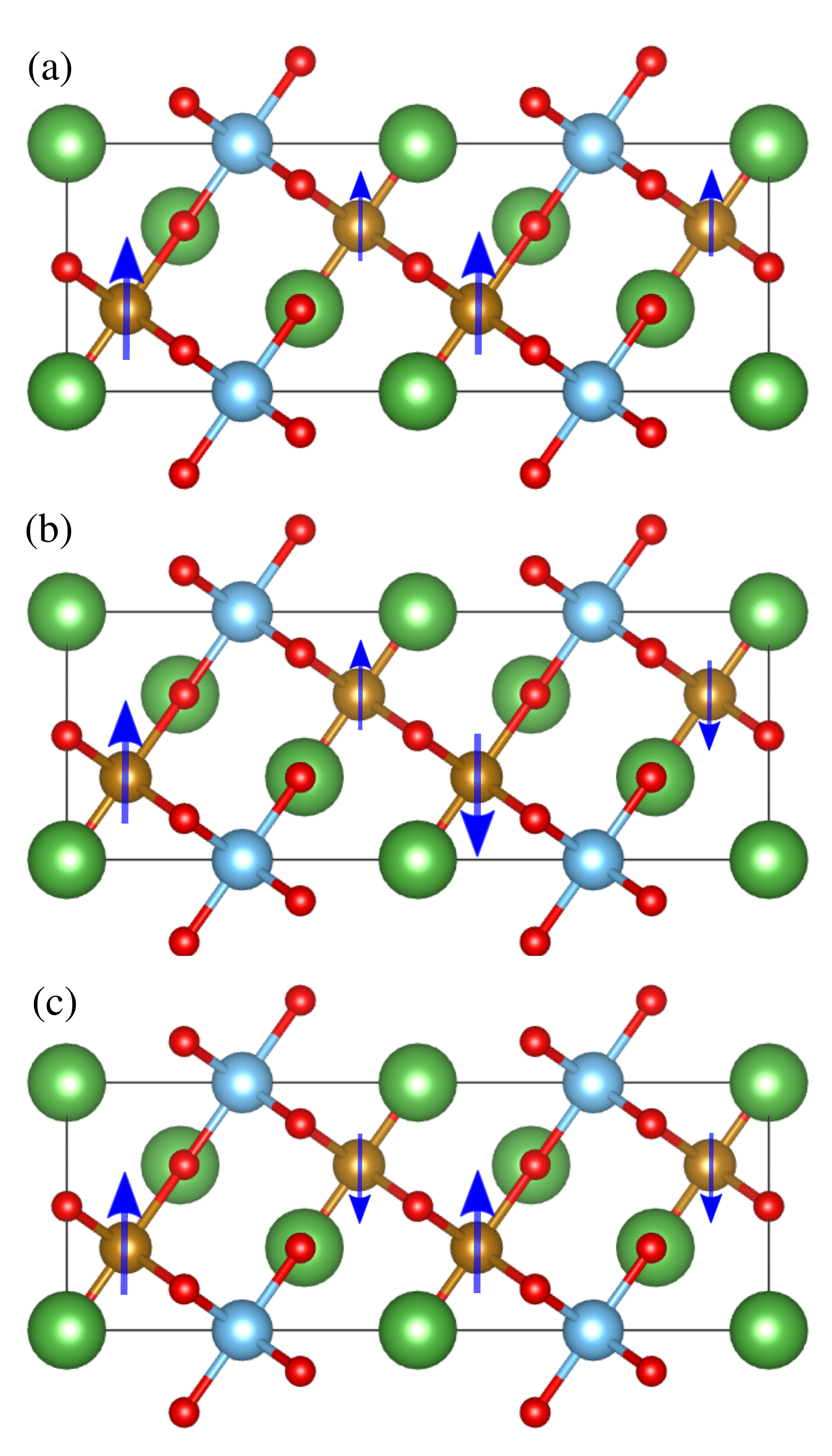}
  \caption{The green, blue, brown, and red spheres are the La, Ti, Fe, and O ions, respectively. The smaller and larger arrows denotes the total spin of Fe$^{2+}$ and Fe$^{3+}$, respectively. In (a), all the magnetic interaction between Fe$^{2+}$ and Fe$^{3+}$ sites are assumed to be ferromagnetic. In (b), antiferromagnetic interaction is assumed between Fe$^{2+}$ and Fe$^{3+}$ nearest neighboring sites, while ferromagnetic interaction is assumed between Fe$^{2+}$ and Fe$^{3+}$ sites across a LaTiO$_3$ layer. In (c), all the magnetic interaction between Fe$^{2+}$ and Fe$^{3+}$ sites are assumed to be ferromagnetic. }
  \label{fig:spin}
\end{figure}

The energies of the structures (Fig. \ref{fig:spin}) with different assumed spin interaction between Fe$^{2+}$ and Fe$^{3+}$ sites were compared. The structure has the lowest energy is in Fig. \ref{fig:spin} (c), in which the antiferromagnetic interaction between Fe$^{2+}$ and Fe$^{3+}$ sites is assumed. The structures in Fig. \ref{fig:spin} (a) and (b) have energies of 134 meV/f.u. and 195 meV/f.u. larger than the structure in Fig. \ref{fig:spin} (c), respectively. Therefore, the antiferromagnetic interaction between Fe$^{2+}$ and Fe$^{3+}$ sites is energetically favorable.

\end{document}